\documentstyle[11pt,newpasp,twoside,epsfig]{article}
\def\edcomment#1{\iffalse\marginpar{\raggedright\sl#1\/}\else\relax\fi}
\reversemarginpar

\begin{document}
\title{Chemical Enrichment and Energetics of the ICM with Redshift}
 \author{Antonio Pipino}
\affil{Dipartimento di Astronomia, Universita' di Trieste, Via G.B. Tiepolo, 
11, 34100 Trieste, Italy}
\author{Francesca Matteucci}
\affil{Dipartimento di Astronomia, Universita' di Trieste, Via G.B. Tiepolo, 
11, 34100 Trieste, Italy}
 \author{Stefano Borgani}
\affil{INFN, Sezione di Trieste, c/o Dipartimento di Astronomia, Universita' di Trieste, Via G.B. Tiepolo, 
11, 34100 Trieste, Italy}
 \author{Andrea Biviano}
\affil{Osservatorio Astronomico di Trieste, Via G.B. Tiepolo, 
11, 34100 Trieste, Italy}

\begin{abstract}
In this paper we show preliminary results concerning the
chemical and energetic enrichment of the ICM by means of 
supernova-driven wind models in
elliptical galaxies.
These models are obtained by taking into account new prescriptions about
supernova remnant evolution in the interstellar medium. 
We find that models, which can reproduce the Fe abundance and 
the [$\alpha$/Fe]
ratios observed in the ICM, predict that the SN energy input
can provide about 0.3 keV per ICM particle.
We have obtained this result by assuming that each SN explosion
inject on the average
into the ISM
no more than 20\% of its initial blast wave energy. The predicted energy 
per particle is not enough to break the cluster self-similarity but is 
more than predicted in previous models.  
\end{abstract}

\section{Introduction}
Hydrodynamical simulations and semi-analytical models for clusters of galaxies need
an extra-energy of about 1 keV per ICM particle (e.g. Wu et al. 2000; Bower et
al. 2000; Bialek et al. 2000; Borgani et al. 2001)
to be reconciled with X-ray observations. In this framework
one of the open problems is to assess whether SNe can provide this
energy in order to break the cluster self-similarity 
(e.g. Finoguenov et al. 2001)
or not (e.g. Valageas \& Silk 1999; Bower et al. 2000). On the other
hand, as shown by Matteucci (this conference), several successfull
models for the chemical enrichment of the ICM have been developed in last
twenty years. We remind that a good model for the ICM chemical 
enrichment must produce
$(Fe/H)/(Fe/H)_{\odot}\sim 0.3$ (e.g. White 2000) and $[\alpha /Fe]\sim 0$
(e.g. Ishimaru \& Arimoto 1997; Renzini 2000).  Since the 
galactic ejecta carry out energy with them we adopted the approach by Matteucci \&
Vettolani (1988) for computing the wind energy in
models which match the observational constraints on the chemistry of the
ICM. We modified the chemical evolution codes developed by Matteucci \& Gibson (1995) and Martinelli et al. (2000) including a new cooling time
for supernova remnants (SNR) depending on metallicity 
(Cioffi \& Shull 1991) as well as new prescriptions
for SNIa cooling (Recchi et al. 2001, section 2). Then we followed 
the evolution of the ICM abundances
with redshift (section 3). In section
4 some results are shown and some conclusions are drawn.

\section{Chemical Evolution Model}
Chemical evolution models (both one and multizone) are those 
of Matteucci \& Gibson (1995) and Martinelli et al. (1998) . We focus here
on the new energetic prescriptions implemented in the chemical evolution code.
We follow the SNR evolution in the interstellar medium according 
to the cooling time by Cioffi \& Shull (1991): 
\begin{equation}
t_{cool} =1.49\cdot10^4 \epsilon_0^{3/14}\,n_0^{-4/7}\,\zeta^{-5/14}\, yr\, ,
\end{equation}
where $\zeta = Z/Z_{\odot}$, $n_0$ is the hydrogen number density,
$\epsilon_0$ is the energy released during a SN explosion in units of
$10^{51}\rm erg$ and we take always $\epsilon_0 =1$. Old (Cox 1972,
long dashed \& dotted line) and new (Cioffi \& Shull 1991) cooling times are compared
in figure 1, where metallicity and density of the ISM evolve in a
self-consistent way as a functions of time. The new cooling time 
(solid line)
is about 3 times lower than the older one after 0.1 Gyr from the
beginning of galactic
evolution and that, soon after 0.2 Gyr, the metallicity becomes
oversolar and makes the cooling $\sim 10$ times more efficient.  The
most important consequence is that, while in previous works flat initial 
mass functions (IMF) were
preferred to get the right amount of elements in the ICM, now we
show that models with Salpeter IMF as the best ones. The reason is that
the faster the metallicity grows the more efficient the cooling
is. So, if we compare a galaxy with a flat IMF (x=0.95) with a galaxy of 
the same initial mass with Salpeter IMF,
the latter undergoes galactic wind earlier than the former and consequently eject more metals into the ICM.
 
\begin{figure}
\centering
\epsfig{file=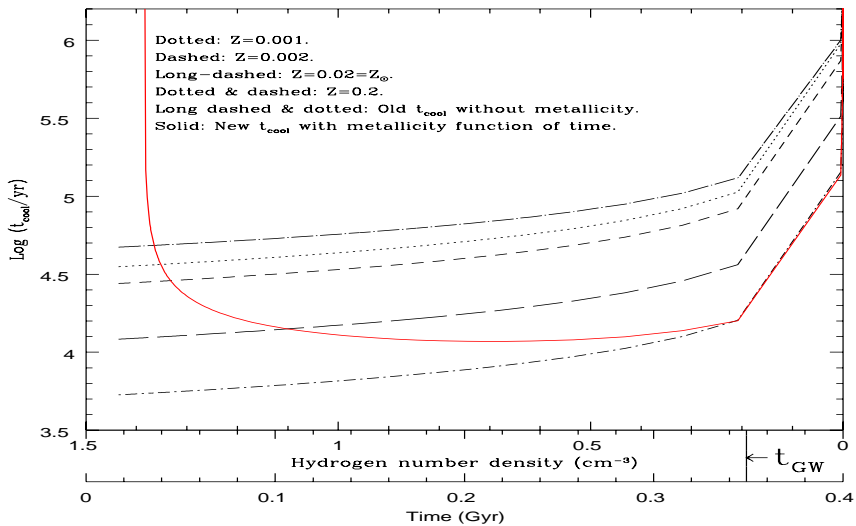, height=7.5cm,width=12cm}
\caption[]{The comparison between different cooling times as a function
of ISM density and metallicity. The new cooling time of Cioffi \& Shull (1991)
is shown by the solid line. The new cooling times at fixed Z are shown with 
dashed and dotted lines. The old cooling time indipendent of Z is indicated by
the long dashed \& dotted line.}
\end{figure}

The second fundamental hypothesis on energetics of the ISM is that 
SNe Ia are allowed to transfer all of their initial blast wave energy 
into the ISM:
\begin{equation}
\epsilon_{SNIa}=1\, .
\end{equation}

In fact Recchi et al. (2001) showed that radiative losses for SNIa are 
negligible, since
their explosions occur in a medium already heated by SNII.
Therefore for SNe II, which explode first in a cold and dense 
medium, we allow for the cooling as described before. This results into 
an efficiency of energy tranfer of
no more than 3\% per SNII. On the other hand for type Ia SNe, exploding only after at least 30-40 Myr, 
we assume an efficiency of energy transfer of 100\%. 

Among all the models we run, here we will present the results of the best ones:

\bf Model MG \rm : best model of Matteucci \& Gibson (1995). It is a one-zone
model with Arimoto \& Yoshii (1987, AY) IMF, cooling time and SNR evolution by Cox (1972)
and no morphological evolution of S0 galaxies allowed. We run 
it for comparison with our models with new energetic prescriptions.

\bf Model I \rm : one-zone model with Salpeter (1955) IMF, Cioffi \& Shull (1991)
cooling time, SNIa without cooling and morphological evolution of S0 galaxies allowed.

\bf Model II \rm : multi-zone model (see Martinelli et al. 2000 for details) with the same
prescriptions of model I.

Despite of the hypothesis on type Ia SNR,
the mean efficiency per SN (Ia+II) explosion is no more than  $\sim 20\%$ for models
I and II while it is $\sim 1.7\%$ for model MG.
Another difference is that, thanks to the new prescriptions, the energy provided by SNIa 
makes galactic winds continuous out to the present time. Therefore models I and II release larger
masses of elements and energy into the ICM than in the MG case. 

\section{Time evolution of Abundances and Energy}

We consider galaxies in the range $10^9 - 2\cdot 10^{12}M_{\odot}$
in order to find the parameters linking the luminous mass $M_l (z)$ of the galaxy
and the masses of the chemical elements as well as the thermal energy $E_{th} (z)$ ejected into the ICM. In particular, we obtain relations of th type:
\begin{equation}
E_{th} (z)=A(z)\, M_{l}^{\beta (z)}(z)\, ,
\end{equation}
\begin{equation}
M_{el} (z)=B(z)\, M_{l}^{\delta (z)}(z)\, ,
\end{equation}
where \it z \rm is the redshift and $A\, , B\, , \beta\, \rm and\it \, \delta$ are least-square fit parameters. For this choice of cosmological paramaters
$\Omega_{m}$=0.3, $\Omega_{\Lambda}$=0.7, $H_{0}$=70 km
$s^{-1}$$Mpc^{-1}$, we integrate relations (3) and (4) over the K-band
Luminosity Function in clusters (LF), taking into account the LF evolution with
redshift (while Martinelli et al. 2000 did not) and the possibility of
morphological evolution for S0 galaxies into spirals for $z \ge 0.4$,
for different cluster richness ($n^*$). In order to do this we
take the LF at z=0 in the B-band from the observational data of Sandage et
al. (1985), then we use the transformation from B-band to K-band by
Fioc \& Rocca-Volmerange (1999) and the evolutionary corrections from
Poggianti (1997). We consider M/$L_K \sim 1$ at z=0 (e.g. Mobasher et
al. 1999), and its evolution in time is calculated by means of the
model of Jimenez et al. (1998). In
order to compare model predictions with observed data we transform the
cluster richness into temperature using the relation $kT \propto
(n^*)^{0.8-1}$.
The choice of the K luminosity is due to the fact that it does not vary 
dramatically with galaxy evolution as it is the case for B luminosity, which 
is very sensitive to young and massive stars.

\section{Results and Conclusions}
In Figure 2 we show the iron abundance (relative to the solar value) predicted
by model II compared to data by White (2000). An AY IMF (solid line)
can lead to a larger amount of Fe in the ICM than the observed value. We
would have seen the same trend if we had used model I, since one and
multi-zone models give quite similar results (as shown also by Table
1). Our models are in agreement also with $[\alpha /Fe]$, but we
predict a slighty lower O abundance.  As we can see from Table 1, SNe
can provide $\sim$ 0.2-0.3 keV per particle. From the chemical
point of view, we show in Figure 3 the evolution of [O/Fe] 
as a function of redshift and metallicity.  At high redshift there is a
fast decrease of O abundance due to the large production of Fe by SNIa,
whereas there is no evidence of evolution for $z\le 1$, in agreement
with observations.
\begin{figure}
\centering
\epsfig{file=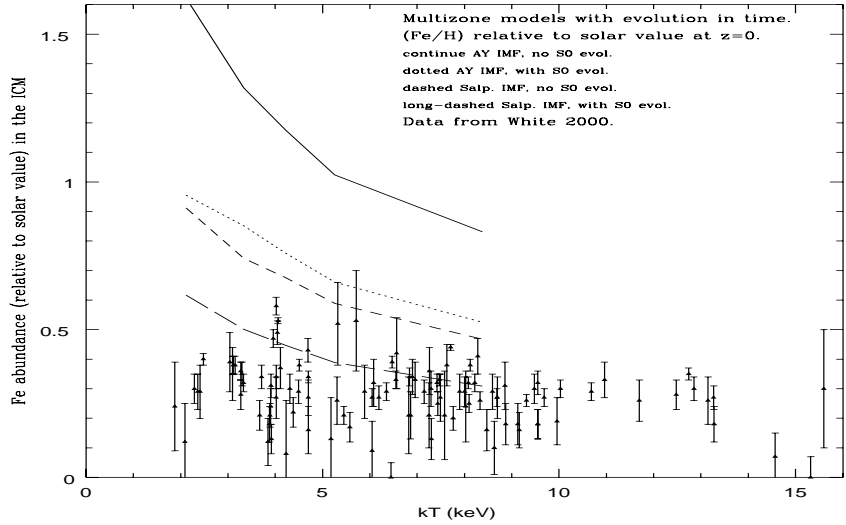, height=7.5cm,width=12cm}
\caption[]{Fe abundances in the ICM as predicted by multizone model compared to
the observed one by White (2000), as a function of cluster temperature.}
\end{figure}

\begin{table}
\small
\begin{flushleft}
\caption[]{Abundances in the ICM and energy per ICM particle for best models with evolution in time for
Coma and Virgo clusters.}
\begin{tabular}{l|lll|ll}
\noalign{\smallskip}
\hline
\hline
\noalign{\smallskip}
Coma  & [O/Fe] obs.& [Si/Fe] obs. &  ${Fe \over Fe_{\odot}}$ obs. \\
\noalign{\smallskip}
\hline
\noalign{\smallskip}
       &$<0.01\pm 0.14>$* &0.51$\pm$ 0.60**   &$<0.23>\dag$ \\
       &                  &$<0.14\pm 0.10>$*     &0.33$\pm$0.05$\ddag$\\ 
\noalign{\smallskip}
\hline
\noalign{\smallskip}
   & [O/Fe]& [Si/Fe]&  ${Fe \over Fe_{\odot}}$& $M_{gas}^{ej}$&$E_{pp}$\\ 
 & & & & $10^{13}M_{\odot}$&({keV})\\  
\noalign{\smallskip}
\hline
\noalign{\smallskip}
MG   &0.09 & 0.21 & $\sim$0.31& 1.27&0.19\\
I &-0.38&    0.003& $\sim$0.24&0.32&0.13\\
II   &-0.66&  -0.08&    $\sim$0.39&0.30&0.22\\
\noalign{\smallskip}
\hline
\hline
\noalign{\smallskip}
Virgo  & [O/Fe] obs.& [Si/Fe] obs.&  ${Fe \over Fe_{\odot}}$ obs. \\
\noalign{\smallskip}
\hline
\noalign{\smallskip}
       &$<0.01\pm 0.14>$* &0.16$\pm$ 0.18**   &0.40$\pm 0.02\P$ \\
       &                  &$<0.14\pm 0.10>$*   &0.55$\pm$0.04$\ddag$\\   
\noalign{\smallskip}
\hline
\noalign{\smallskip}
   & [O/Fe]& [Si/Fe]&  ${Fe \over Fe_{\odot}}$& $M_{gas}^{ej}$&$E_{pp}$\\ 
 & & & & $10^{13}M_{\odot}$&({keV})\\  
\noalign{\smallskip}
\hline
\noalign{\smallskip}
MG   &0.08 & 0.21 & $\sim$0.50&  0.51&0.30\\
I &-0.38&      0.003&  $\sim$0.35& 0.13&0.21\\
II   &-0.66&      -0.08&	 $\sim$0.61 &0.12&0.34\\
\noalign{\smallskip}
\hline
\hline
\noalign{\smallskip}

\end{tabular}
\end{flushleft}
 $\dag$ (Fe/H)observed in Coma cluster from De Grandi \& Molendi (2001, Beppo-Sax);$\ddag$ by Matsumoto et al. (2000).
$\P$(Fe/H)observed in Virgo cluster from White (2000, ASCA).
**[Si/Fe] from Fukazawa et al. (1998).
*[Si/Fe], [O/Fe] weighted mean from Ishimaru \& Arimoto (1997) for cluster A496, A1060, A2199 and AWM7.
$M_{gas}^{ej} \sim 1-20 \%$ $M_{ICM}$.
\end{table}
\begin{figure}
\centering
\epsfig{file=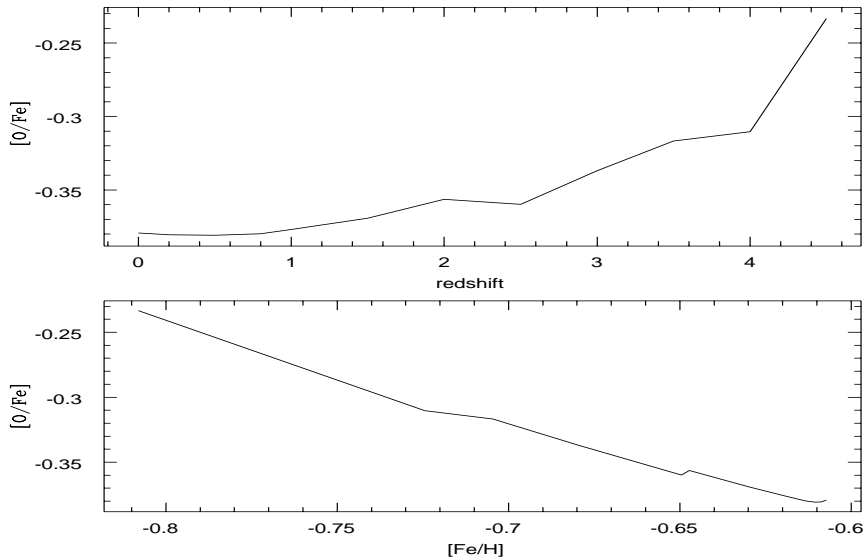, height=8cm,width=12cm}
\caption[]{Evolution of [O/Fe] versus redshift in Coma-like cluster.}
\end{figure}

In summary, our conclusions are:
\begin{itemize} 
\item In both one and multi-zone models, only those with Salpeter IMF
     reproduce the ICM Fe abundance as well as the observed [Si/Fe] ratios.
\item Best models can provide 0.2-0.3 keV  per ICM particle
      with a mean SN efficiency of $\sim 20\%$.	
\item No relevant evolution is found for abundances and energy per particle from
      z=0 out to z=1, in agreement with observational data (e.g. Matsumoto et al. 2000).
\item In both old and new models type Ia SNe play a fundamental role in
      providing energy ($\sim 80-95 \% $) and Fe ($\sim 45-80 \% $) 
      to the ICM.
\item In spite of the large contribution from SNIa, SNe in general
      seem not to be able to provide the requested 1 keV per particle.
      We need perhaps other energy sources 
      (e.g. quasars), although we might have underestimated the energy 
      contribution from type II SNe.
\item Chandra and XMM observations on abundances and abundance gradients
      in the ICM will set stronger constraints to our models. 
\end{itemize}

\end{document}